\documentstyle{mn}
\input psfig.tex
\newif\ifAMStwofonts

\ifoldfss
  \ifCUPmtlplainloaded \else
    \NewTextAlphabet{textbfit} {cmbxti10} {}
    \NewTextAlphabet{textbfss} {cmssbx10} {}
    \NewMathAlphabet{mathbfit} {cmbxti10} {} 
    \NewMathAlphabet{mathbfss} {cmssbx10} {} 
  \fi
  \ifAMStwofonts
    \ifCUPmtlplainloaded \else
      \NewSymbolFont{upmath} {eurm10}
      \NewSymbolFont{AMSa} {msam10}
      \NewMathSymbol{\upi}     {0}{upmath}{19}
      \NewMathSymbol{\umu}     {0}{upmath}{16}
      \NewMathSymbol{\upartial}{0}{upmath}{40}
      \NewMathSymbol{\leqslant}{3}{AMSa}{36}
      \NewMathSymbol{\geqslant}{3}{AMSa}{3E}

      \let\geq=\geqslant 
    \fi
  \fi
\fi 
\ifnfssone
  \newmathalphabet{\mathit}
  \addtoversion{normal}{\mathit}{cmr}{m}{it}
  \addtoversion{bold}{\mathit}{cmr}{bx}{it}
  \newmathalphabet{\mathbfit} 
  \addtoversion{normal}{\mathbfit}{cmr}{bx}{it}
  \addtoversion{bold}{\mathbfit}{cmr}{bx}{it}
  \newmathalphabet{\mathbfss} 
  \addtoversion{normal}{\mathbfss}{cmss}{bx}{n}
  \addtoversion{bold}{\mathbfss}{cmss}{bx}{n}
  \ifAMStwofonts
    \ifCUPmtlplainloaded \else
      \UseAMStwoboldmath
      \makeatletter
      \new@mathgroup\upmath@group
      \define@mathgroup\mv@normal\upmath@group{eur}{m}{n}
      \define@mathgroup\mv@bold\upmath@group{eur}{b}{n}
      \edef\UPM{\hexnumber\upmath@group}
      \new@mathgroup\amsa@group
      \define@mathgroup\mv@normal\amsa@group{msa}{m}{n}
      \define@mathgroup\mv@bold\amsa@group{msa}{m}{n}
      \edef\AMSa{\hexnumber\amsa@group}
      \makeatother
      \mathchardef\upi="0\UPM19
      \mathchardef\umu="0\UPM16
      \mathchardef\upartial="0\UPM40
      \mathchardef\leqslant="3\AMSa36
      \mathchardef\geqslant="3\AMSa3E

      \let\geq=\geqslant 
    \fi
  \fi
\fi 

\ifnfsstwo
  \DeclareMathAlphabet{\mathbfit}{OT1}{cmr}{bx}{it}
  \SetMathAlphabet\mathbfit{bold}{OT1}{cmr}{bx}{it}
  \DeclareMathAlphabet{\mathbfss}{OT1}{cmss}{bx}{n}
  \SetMathAlphabet\mathbfss{bold}{OT1}{cmss}{bx}{n}
  \ifAMStwofonts
    \ifCUPmtlplainloaded \else
      \DeclareSymbolFont{UPM}{U}{eur}{m}{n}
      \SetSymbolFont{UPM}{bold}{U}{eur}{b}{n}
      \DeclareSymbolFont{AMSa}{U}{msa}{m}{n}
      \DeclareMathSymbol{\upi}{0}{UPM}{"19}
      \DeclareMathSymbol{\umu}{0}{UPM}{"16}
      \DeclareMathSymbol{\upartial}{0}{UPM}{"40}
      \DeclareMathSymbol{\leqslant}{3}{AMSa}{"36}
      \DeclareMathSymbol{\geqslant}{3}{AMSa}{"3E}

      \let\geq=\geqslant 
    \fi
  \fi
\fi 

\ifCUPmtlplainloaded \else
  \ifAMStwofonts \else 
    \def\upi{\pi}
    \def\umu{\mu}
    \def\upartial{\partial}
  \fi
\fi

\title{Global Solutions of Viscous Transonic Flows in Kerr Geometry I:
Weak Viscosity Limit}
\author[Sandip K. Chakrabarti]
       {Sandip K. Chakrabarti \\
Tata Institute Of Fundamental Research, Homi Bhabha Road, Mumbai 400005, INDIA}
\date{Accepted .
      Received ;
      in original form }
\pubyear{1996}

\begin{document}

\maketitle

\begin{abstract}

We present fully general relativistic equations governing 
viscous transonic flows in vertical equilibrium in Kerr 
geometry. We find the complete set of global solutions 
(both for Optically thick and optically thin flows) in 
the weak viscosity limit. We show that for a large region of 
parameter space, centrifugal pressure supported standing shocks 
can form in accretion and winds very close to the black hole
horizon, both for co-rotating and contra-rotating flows. We 
compute the nature of the shear tensor for complete transonic 
solutions and discuss the consequences of its reversal properties.

\end{abstract}


\noindent Appeared in MNRAS v. 283, p. 325, 1996

\section{Introduction}

The physics of accretion processes on black holes has fascinated astrophysicists
for last two decades (Novikov \& Thorne 1973, hereafter NT73; 
Lynden-Bell, 1978; Moncrief, 1980) but the full complexity of the 
flow behaviour is yet to be unfolded in a comprehensive manner. For example,
one of the aspects that is recently studied more actively, is the accretion flow with
`discontinuities' or shock waves. The important numerical works of Wilson (1978) and
Hawley, Smarr \& Wilson (1984, 1985) show that sub-Keplerian flows experience
centrifugal barrier close to the black hole and produce shock waves.
Though these fully general relativistic
simulations revealed the existence of only traveling 
shock waves, subsequent theoretical and numerical study using {\it pseudo-Newtonian 
potentials} show that shocks are indeed stationary in a large region of the
parameter space (Chakrabarti, 1989 [C89]; 1990a,b [C90a],[C90b]; 
Molteni, Lanzafame \& Chakrabarti, 1994; Sponholz \& Molteni, 1994). Similarly,
the formation of sub-Keplerian flows from Keplerian disks in radiation 
(Paczy\'nski \& Wiita, 1980) and ion pressure (Rees et al., 1982) supported disks
have been shown only in the context of pseudo-Newtonian geometry
(C90ab, Chakrabarti, 1996 [C96]).
So far, no global solution of rotating, viscous, transonic flow
(with cooling effects included) has been obtained using full
general relativistic considerations. Only inviscid solutions
of conical flows (with simplistic accretion rate equation) are
present (C90b).

In the present paper, we derive the equations governing
the viscous transonic flows in Kerr geometry and find global 
solutions using the usual sonic point analysis. We present explicit examples 
of solutions in weak viscosity limit. We see that similar to our earlier
studies using peudo-potentials, standing shocks can form
around Kerr black holes for a large range of initial flow parameters.
The maximum number of sonic points remains three as was first pointed out
by Liang and Thompson (1980) in the context of equatorial accretion
in Schwarzschild geometry.
The important difference between the present study and the earlier
works (e.g., Chakrabarti 1989) is that the present work
does not use pseudo-potential and therefore clearly satisfies the
boundary condition on the horizon. Secondly, shock locations,
particularly in prograde flows,
are very close to the black hole, roughly at half as much
distance as those obtained in Schwarzschild geometry. In 
retrograde flows, the shocks form farther away. These may have
important implications in the study of hard radiations from black hole
candidates (e.g., Chakrabarti \& Titarchuk, 1995). Also, for the first time,
the nature of shear and angular momentum transport in a
viscous transonic flow just outside the black hole horizon are discussed.

In the next Section, we present the basic equations. In \S 3, we solve
these equations in weak viscosity limit and show examples of global 
solutions with or without stationary shocks. In \S 4, we present
the behavior of shear tensor which is responsible for 
angular momentum transport. Finally, in \S 5, we make concluding remarks.

\section{Basic Equations}

We use $t$, $r$, $\theta$ and $z$ as the coordinates. We choose the geometric 
units where $G=M_{bh}=c=1$ ($G$ is the gravitational constant,
$M_{bh}$ is the mass of a black hole and $c$ is the velocity of light). 
We also consider $|\theta-\pi/2| <<1 $
for a thin flow in vertical equilibrium.
We consider a perfect fluid with the stress-energy tensor,
$$
T_{\mu\nu} = \rho u_\mu u_\nu + p(g_{\mu\nu} + u_\mu u_\nu)
\eqno{(1)}
$$
where, $p$ is the pressure and $\rho=\rho_0(1+\pi)$ is the 
mass density, $\pi$ being the internal energy.
We ignore the self-gravity of the flow.
We assume the vacuum metric around a Kerr black hole to be of the form (e.g., NT73),
$$
ds^2 = g_{\mu\nu}dx^\mu dx^\nu= -\frac{r^2 \Delta}{A} dt^2 + \frac{A}{r^2}
(d\phi-\omega dt)^2 +\frac{r^2}{\Delta} dr^2 + dz^2
\eqno{(2)}
$$
Where,
$$
A= r^4 + r^2 a^2 + 2 r a^2
$$
$$
\Delta= r^2 - 2 r + a^2
$$
$$
\omega = \frac{2 a r}{A}
$$
Here,  $g_{\mu\nu}$ is the metric coefficient and 
$u_\mu$ is the four velocity components:
$$
u_t= \left[\frac{\Delta}{(1-V^2)(1-\Omega l)(g_{\phi\phi}+l g_{t\phi})}\right]^{1/2}
\eqno{(3a)}
$$
and
$$
u_\phi=-l u_t
\eqno{(3b)}
$$
where, the angular velocity is
$$
\Omega=\frac{u^\phi}{u^t}=-\frac{g_{t\phi}+ l g_{tt}}{g_{\phi\phi}+l g_{t\phi}}
\eqno{(4)}
$$
and $l$ is the specific angular momentum. The radial velocity $V$ in the rotating frame is
(C90b, p. 137)
$$
V=\frac{v}{(1-\Omega l)^{1/2}}
$$
where,
$$
v=(-\frac{u_ru^r}{u_tu^t})^{1/2}.
$$
Since $V=1$ on the horizon and
even in the extreme equation of state of $p=\rho/3$, the flow must 
be supersonic on the horizon. Thus, {\it any black hole accretion is necessarily transonic}
(also, see C90b). 

\subsection{Equations Governing a Viscous Transonic Flow}

In the present paper, we shall concentrate on the time independent 
solutions of the underlying hydrodynamic equations.
The equation for the balance of the radial momentum is obtained from
$(u_\mu u_\nu + g_{\mu\nu}) T^{\mu\nu}_{;\nu}=0$:
$$
\vartheta\frac{d\vartheta}{dr} + \frac{1}{r \Delta} [a^2- r + \frac{A\gamma^2 B}{r^3}]\vartheta^2
+ \frac{A \gamma^2}{r^6}B 
+ (\frac{\Delta}{r^2} + \vartheta^2 )\frac{1}{p+\rho} \frac{dp}{dr}=0
\eqno{(5)}
$$
where,
$$
\gamma^2= [1-\frac{A^2}{\Delta r^4} (\Omega - \omega)]^{-1} ,
$$
$$
B=(\Omega a - 1)^2 - \Omega^2 r^3 ,
$$
and
$$
\vartheta=u^r.
$$
Here and hereafter we use a comma to denote an ordinary derivative and a 
semi-colon to denote a covariant derivative.
The baryon number conservation equation (continuity equation)
is obtained from $(\rho_0 u^\mu)_{; \mu}=0$ which is,
$$
\noindent {\dot M}= 2 \pi r \vartheta \Sigma= 2\pi r \vartheta \rho_0 H_0
\eqno{(6)}
$$
where,
$$
H_0= (\frac{p}{\rho_0})^{1/2} \frac{r^{3/2}}{\gamma} 
[\frac{(r^2+a^2)^2- \Delta a^2}{(r^2 + a^2)^2 + 2 \Delta a^2}]^{1/2}
$$
is the height of the disk in vertical equilibrium (NT73).
The equation of the conservation of angular momentum 
is obtained from $(\delta^\mu_\phi T^{\mu \nu})_{;\nu}=0$, and one obtains,
$$
\rho_0 u^\mu (h u_\phi)_{,\mu}= (\eta \sigma_\phi^\gamma)_{;\gamma}
$$
where,
$$
\eta=\nu\rho_0
$$
is the coefficient of dynamical viscosity and $\nu$ is the coefficient of
kinematic viscosity. When rotation is dominant ($\vartheta <u^\phi$)
the relevant shear tensor component $\sigma_\phi^r$ is  given by (Anderson \& Lemos, 1988),
$$
\sigma_\phi^r= - \frac{A^{3/2} \gamma^3 \Omega_{,r} \Delta^{1/2}}{2r^5}
\eqno{(7)}
$$
so that the angular momentum equation takes the form,
$$
{\cal L}-{\cal L}_{+}=  -\frac{1}{\vartheta r^5}\frac{d\Omega}{dr} \nu A^{3/2} \gamma^3 \Delta^{1/2} .
\eqno{(8)}
$$
Here we have modified earlier work of NT73 and Peitz (1994) in that the fluid
angular momentum ${\cal L}=-h u_\phi$, rather than particle angular
momentum $-u_\phi$ is used. That way, for an inviscid flow ($\eta=0$)
one recovers ${\cal L}=constant$ as in a fluid picture. Similarly, the radial velocity term is 
included (eq. 3) and angular momentum is allowed to be non-Keplerian (eq. 8).
${\cal L}_+$ is the angular momentum on the horizon since the {\it rotational} shear (as defined
by eq. 7) vanishes there.
In presence of significant radial velocity, the shear in eq. (7)
is to be replaced by its full expression, $\sigma^{\mu\nu}=
(u^\mu_{;\beta} P^{\beta\nu} + u^{\nu}_{;\beta}P^{\beta\mu})/2 - \Theta P^{\mu\nu}/3$
where $P^{\mu\nu}= g^{\mu\nu}+u^\mu u^\nu$ is the projection tensor and $\Theta=
u^\mu_{;\mu}$ is the expansion (e.g., Shapiro \& Teukolsky, 1983) which 
no longer vanishes on the horizon (see, Fig. 6 below).
In that case, ${\cal L}_+$ will no longer
be the specific angular momentum of the flow at the horizon, instead it will be at
the place where the net shear effect 
vanishes. In Section 4, we shall compare these shear components
in a realistic solution.

Entropy generation equation is obtained from the first law of thermodynamics
along with the baryon conservation equation $(S^\mu)_{;\mu}=
[2\eta \sigma_{\mu\nu}\sigma^{\mu\nu}]/T - Q^-$ :
$$
\vartheta \Sigma (\frac{dh}{dr} - \frac{1}{\rho_0}\frac{dp}{dr}) =  Q^+-Q^- =
2\nu \Sigma \sigma_{\mu\nu}\sigma^{\mu\nu} - Q^ -
$$
where $Q^+$ and $Q^-$ are the heat generation rate
and the heat loss rate respectively. $h$ is the specific enthalpy: 
$h=(p+\rho)/\rho_0$. Here, we ignore the terms contributed by radiation.
Using rotational shear as given in eq. (7), the entropy 
equation takes the form,
$$
\vartheta \Sigma (\frac{dh}{dr} - \frac{1}{\rho_0}\frac{dp}{dr}) = 
\frac{\nu \Sigma A^2 \gamma^4 (\Omega_{,r})^2}{r^6} - Q^- .
\eqno{(9)}
$$
Of course, for accuracy, we use the full expression for $\sigma_\phi^r$
as discussed above.

This set of equations are solved simultaneously
keeping in mind that the shock waves may form
in the flow, where, the following momentum balance condition
$$
W_-n^\nu + (W_-+\Sigma_{0-}) (u_-^\mu n_\mu)u_-^\nu  =
W_+n^\nu + (W_++\Sigma_{0+}) (u_+^\mu n_\mu) u_+^\nu
\eqno{(10)}
$$
along with the conservation of energy and mass fluxes (together, these conditions are
known as the Rankine-Hugoniot conditions)  must be fulfilled.
Here, $n_\mu$ is the four normal vector component across the shock, and $W$ and
$\Sigma$ are vertically integrated pressure and density on the shock surface.
The subscripts $-$ and $+$ denote the pre-shock and post-shock quantities
respectively. 

The equations presented above are applicable to optically thin as well as
optically thick flows for any general heating and cooling processes. 
For a given viscosity prescription and the exact cooling 
processes (depending on the optical depth of the flow),
it is usual to reduce the above set of equations in the form (Bondi, 1952):
$$
\frac{du}{dr}=\frac{N}{D}
\eqno{(11)}
$$
where $N$ and $D$ are the smooth functions of radial coordinate (unless
there are non-linearities which prevent such reduction. In that case
sonic curve analysis is done, see Flammang, 1982). The procedure of
obtaining the complete solution is then similar to what is presented
in obtaining the global solutions of viscous transonic flow using
pseudo-Newtonian potential (C90a,b; C96 and references therein).
Presently, we concentrate on the
solution of the above set of equations in the inviscid limit. Full 
discussion on strongly viscous flows will be presented elsewhere.

\subsection{Sonic Point Conditions in Weak Viscosity Limit}

In the case when viscosity is negligible, equations (5) and (8) could be 
integrated to obtain the energy and angular momentum conservation equations
(Moncrief, 1980; Flammang, 1982; C90b),
$$
hu_t = \frac {p+ \rho} {\rho_0} u_t = {\cal E},
\eqno{(12a)}
$$
and
$$
hu_\phi =\frac {p + \rho}{\rho_0} u_\phi = -{\cal L}.
\eqno{(12b)}
$$
From eq. (6) we get the constant accretion rate as,
$$
{\dot M} = 2\pi r \vartheta \rho_0 H_0 .
\eqno{(12c)}
$$

In the weak viscosity limit, the entropy equation (9) can be replaced by an
adiabatic equation of state $P=K \rho_0^{1+1/n}$. $K$, which is a measure of entropy,
is constant in the flow but it changes at the shocks (the
entropy generated may be either radiated or advected away depending on
the cooling effciency of the flow at the shock; see C90b), $n$ is the polytropic 
index related to the adiabatic index $\Gamma$ by $\Gamma=1+1/n$, and is assumed to be
constant throughout the flow. The specific enthalpy
and the density in terms of the sound speed $a_s=(\partial P/\partial \rho)^{1/2}$ become,
$$
h=\frac{1}{1-n{a_s^2}}
\eqno{(13a)}
$$
$$
\rho_0=[\frac{n a_s^2}{K (n+1)(1-na_s^2)}]^n
\eqno{(13b)}
$$

First we rewrite the energy and mass conservation equations
on the equatorial plane as ($\theta=\pi/2$),
$$
{\cal E}= \frac{1}{1-na_s^2} \frac{1}{(1-V^2)^{1/2}} 
\left(\frac{\Delta}{\cal D}\right )^{1/2}
\eqno{(14a)}
$$
and
$$
{\dot {\cal M}} =( \frac{a_s^2}{1-na_s^2})^{n+1/2}\frac{V}{(1-V^2)^{1/2}} 
[\frac{\Delta r^3 \{(r^2+a^2)^2- \Delta a^2\}}
{\gamma\{(r^2 + a^2)^2 + 2 \Delta a^2\} }]^{1/2}
\eqno{(14b)}
$$
where,
$$
{\cal D}= (1-\Omega l)(g_{\phi\phi}+l g_{t\phi})
= r^2 + \frac{2l^2}{r}(1-\frac{a}{l})^2-l^2(1-\frac{a^2}{l^2}) .
\eqno{(15a)}
$$
$$
v_\phi =(-\frac{u_\phi u^\phi} {u_tu^t})^{1/2} .
\eqno{(15b)}
$$
The quantity ${\dot {\cal M}}$ will be
called the {\it entropy accretion rate} as ${\dot {\cal M}} \propto K^n {\dot M}$.
It usually assumes two different values at two saddle type sonic points and 
the one with the smallest ${\dot {\cal M}}$ 
joins horizon with infinity (C89, C90b).
This quantity, which is very useful in the study of shocks,
was defined for the first time in C89 and was simply
termed as `accretion rate'.  In presence of an accretion  shock, the transonic flow
passes through both the sonic points.

Differentiating equations (14a) and (14b), and eliminating
terms involving derivatives of $a_s$, we obtain,
$$
\frac{dV}{dr} = \frac {N}{D}= \frac{c_1 a_2 - c_2 a_1}{b_1 a_2 - b_2 a_1}
\eqno{(16)}
$$
where,
$$
a_1=\frac{2na_s}{1-na_s^2}
\eqno{(17a)}
$$
$$
a_2=\frac{2n+1}{a_s(1-na_s^2)}
\eqno{(17b)}
$$
$$
b_1=\frac{V}{1-V^2}
\eqno{(17c)}
$$
$$
b_2=\frac{1}{V(1-V^2)}
\eqno{(17d)}
$$
$$
c_1=\frac{r-1}{\Delta}-\frac{r-\frac{l^2}{r^2}(1-\frac{a}{l})^2}{\cal D}
\eqno{(17e)}
$$
$$
c_2= \frac{3}{2r}+\frac{r-1}{\Delta} -\frac{\gamma^{'}}{\gamma} + \frac{H_0^{'}}{H_0}
\eqno{(17f)}
$$
Where a prime ($'$) denotes ordinary derivative with respect to r.
From the vanishing condition of $N$ and $D$ at the sonic points,
one obtains the so called sonic point conditions as,
$$
V_c^2=\frac{c_1}{c_2}|_c
\eqno{(18a)}
$$
and
$$
a_{s,c}=V_c(\frac{2n+1}{2n})^{1/2}
\eqno{(18b)}
$$
Here, the subscript `c' refers to the quantities evaluated at the sonic point 
$r=c$. The sonic point condition (7b) implies a Mach number $M=V/a_s=
(2n/2n+1)^{1/2}$ at the sonic point. This is of the same form as the flow in
vertical equilibrium in pseudo-Newtonian potential (C89), though the
velocities here are defined with relativistic corrections. In conical flows
in Kerr geometry (with a simpler accretion rate equation),
$M=1$ at the sonic point (C90b).

In obtaining a global solution one supplies the conserved
quantities at the inner (such as the disk surface) or the outer
(e.g, Keplerian or sub-Keplerian flows injected at the outer region) 
boundary, depending upon whether one 
is interested in the wind solution or the accretion solution. 
For a given angular momentum $l$, the remaining
unknowns are $V (r)$, $a_s (r)$. But one requires
only one extra boundary condition, e.g., ${\cal E}$, since
two sonic point conditions (eqs. 18a and 18b) introduce only one extra
unknown, namely, $r_c$.  Thus, the supply of the initial
specific energy ${\cal E}$ and the specific angular momentum $l$
are sufficient for a complete solution from the horizon to infinity.
For a viscous flow, one clearly has to supply the distribution of
viscosity $\eta(r)$ (e.g., ion or magnetic viscosity) itself. Simple
Shakura \& Sunyaev  (1973) viscosity prescription may not be
very useful since that stress $-\alpha P$ is always negative,
while in a general relativistic flow stress can change signs.
Note that by definition, $\Omega=\omega$ on the horizon
and thus the flow co-rotates with the black hole.
Instead of specifying various quantities at the flow boundary,
one can alternatively specify the location of a critical point
along with the energy or the angular momentum as in C90a,b. 

In much of the parameter space, the flow is expected to be smooth
as in a Bondi flow. However, if the angular momentum is significant,
matter can pile up behind the  centrifugal barrier close to the
black hole and form a standing shock wave. 
At the shock, apart from the continuity of the energy and mass flux,
the relativistic momentum balance condition (eq. 10)
must be satisfied. Using Newtonian definition of the vertical integration
(since we are dealing with thin flows anyway) as in C89 and the definition
of entropy accretion rate ${\dot {\cal M}}$, we find that at the shock, the
following quantity:
$$
\Pi = \frac{\left [ \frac{a_{s}^2}{1-na_{s}^2} \right ]^{n+3/2}
\left ( \frac{2}{3\Gamma-1} + \frac {V^2}{a_s^2 (1-V^2)} \right )}
{{\dot{\cal M}}} 
\eqno{(19)}
$$
should be continuous. One can use different prescriptions, but the 
general behaviour is not expected to change.
 
\section{Behaviour of Sonic Points and Global Solutions}

Before obtaining a globally complete solution, we can make a general study
of the sonic point behaviour of a weakly viscous flow.
Using eq. 14a, eq. 18a and eq. 18b, we can compute the specific 
energy ${\cal E}$  of the flow (which is conserved)
as a function of the sonic radius $r_c$ and specific angular
momentum of the flow. Figs. 1(a-b) shows our findings for $a=0.95$.
We choose $\Gamma=4/3$, i.e., $n=3$.
In Fig. 1a, from the top curve to the bottom we use,
$l=2.0, \ 2.1,\ 2.2,\ 2.3,\ 2.4$ and $ 2.5$.
The number of sonic points of the flow for a given energy and angular momentum
will depend on the number of intersections between a ${\cal E}=$ constant line
and a constant angular momentum curve. If the slope of the curve
is negative the sonic point is saddle type and is physical, 
otherwise it is of circle type or `O' type (see, C89, C90a,b) which is unphysical
and irrelevant. We note that for a very high energy ${\cal E}>>1$ or for
low angular momentum there is only one saddle type sonic point as in Bondi accretion
(Bondi, 1952). But near ${\cal E}=1$ (i.e., the rest mass), 
there could be three sonic points (see Fig. 3 below).
The innermost and the outermost intersections represent saddle
type sonic points and the middle intersection represents the `O' type
sonic point. For low angular momentum (say, for $l=2.0$) there is
only one sonic point for all energy with which matter enters the black hole. This is
similar to a spherical Bondi flow. However, for higher angular
momentum, say $l=2.3$, there could be two saddle type sonic points for
a range of energy. The presence of the second saddle type 
point is the general relativistic effect, as discussed in C90b or C96.
In Fig. 1b, we present results for retrograde flows with angular momenta
(from the top  curve to the bottom), $l= 3.4,\ 3.8,\ 4.2,\ 4.6,\ 5.0$ and $5.4$.
Here, to have three sonic points the flow requires a larger angular 
momentum. This is because the
matter has to fight against the frame-dragging effects which 
forces the flow to co-rotate with the horizon. The terms containing $a/l$ 
in Eq. (15a) and Eq. (17e) represent the coupling  of the spin angular
momentum of the black hole and the orbital angular momentum of the
particle which bring about this change of behaviour. In either cases,
however, the flow angular momentum that we consider here are around
marginally stable value or even less. Thus, they are not high in any real sense.

For a higher polytropic index, say, for $\Gamma=5/3$, the number of 
saddle type sonic points is only one. The $\Gamma=\Gamma_{crit}$
at which the transition from two saddle type points to just one saddle type
sonic point takes place depends on the geometry of the flow.
Typically, we find that for a flow in vertical equilibrium, $\Gamma_{crit}\sim 1.5$
(see also, C90b; C96). 

\begin{figure}
\psfig{figure=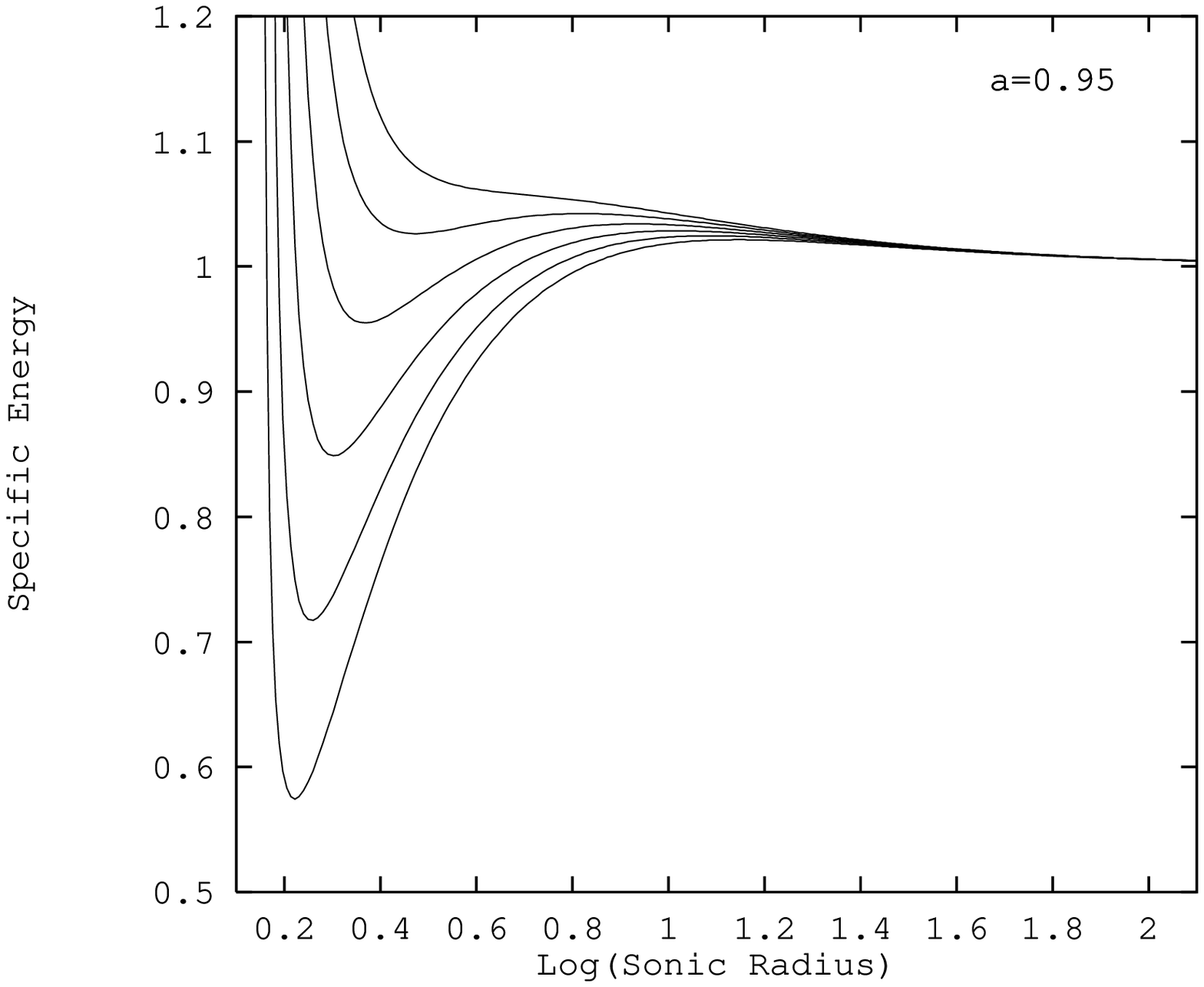,height=8.0truecm,width=8.0truecm,angle=270}
\noindent {\small {\bf Fig. 1a:} Plot of specific energy 
${\cal E}$ as functions of
the sonic point ($x_c$) for $l=2.0,\ 2.1,\ 2.2,\ 2.3,\ 2.4$ and $2.5$
(from top to bottom). $a=0.95$ is used. For some range of energy, the curves
of certain $l$ may have three intersections, signifying the presence
of three sonic points and possibility of shock waves.}
\end{figure}

\begin{figure}
\psfig{figure=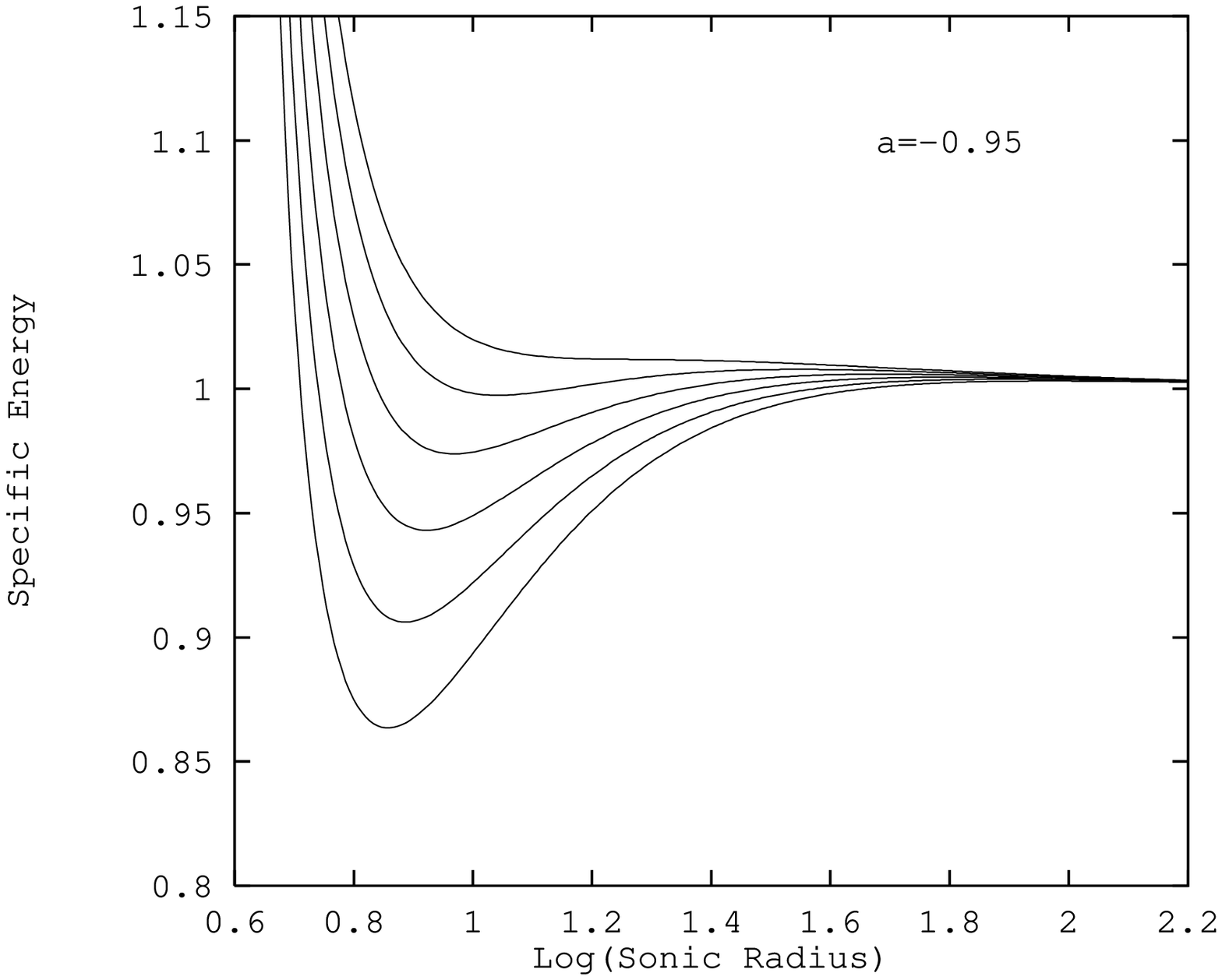,height=8.0truecm,width=8.0truecm,angle=270}
\noindent {\small {\bf Fig. 1b:} Same as (a) but for contra-rotating flow with
$l= 3.4,\ 3.8,\ 4.2,\ 4.6,\ 5.0$ and $5.4$ (from top to bottom).
The sonic points are located farther out in this case. }
\end{figure}

Another important property of the flow which could be studied
before solving the full set of equations is the dependence of
entropy accretion rate on specific energy at various sonic points 
for a given angular momentum of the flow. Since one requires only two 
free parameters for a global solution, when a pair (${\cal E}$, $l$) 
is supplied at the outer boundary, ${\dot{\cal M}}$ must be regarded as an
eigenvalue of the problem. 
Note that only the product of entropy function $K^n$ and the mass accretion rate
${\dot M}$ is determined by the boundary condition, not the
individual quantities. Thus, a specific flow property,
such as shock location, could be completely independent of the mass 
flux ${\dot M}$. When cooling processes (which depend on mass accretion rate) 
are used, the result will depend on accretion rates, except when
cooling relation is simple (such as $Q^-/Q^+=$constant, see C96).

In Fig. 2, we present the variation of the specific energy ${\cal E}$
with entropy accretion rate ${\dot {\cal M}}$ as the sonic point
is changed. This is for inviscid flow ($\eta(r)=0$). 
The direction of the arrow on the curve indicates the direction 
in which the sonic point is increased. The specific angular momentum 
of the flow is $l=2.3$ and $a=0.95$ is chosen as before. 
The branches $AMB$ and $CMD$ are drawn for the inner 
and the outer sonic points respectively and the branch $BC$ is drawn 
for the circle type sonic point (C89). For a given mass accretion rate
${\dot M}$, the entropy accretion rate measures the specific entropy
of the flow. In the energy range spanned by $MC$, 
${\dot {\cal M}}_o > {\dot {\cal M}}_ i$, 
and in the energy range spanned by $MD$,
${\dot {\cal M}}_o < {\dot {\cal M}}_ i$. (Here, the subscripts
$i$ and $o$ denote the quantities at the inner and the outer sonic
points respectively.) However, since only smaller
of the two ${\dot {\cal M}}$s joins the flow at infinity (C90b), only allowed
(shock free) flow must have parameters either on $AM$ (accretion
or wind through inner sonic point) or on $MD$ 
(accretion or wind through outer sonic point). In some regions
of the parameter space close to $M$, two different transonic flows
(e.g., two flows with the same energy and angular momentum, but different
entropy accretion rate) can be `glued' together through shock conditions. 
For example, a flow, first passing through an outer sonic point on $MD$
can undergo a shock $aa$ and generate enough entropy to `land on' the branch
$MB$ so that it can now pass through the inner sonic point. This
is an example of the accretion shock. We shall show this solution
in Fig. 3 below. Conversely, a wind flow can first pass through an inner sonic
point on $MA$ and after passing through a shock can land on a point on 
the branch $MC$ so that it can pass through the outer sonic point also.
This solution is also shown in Fig. 3 below.
In order to have a shock, of course, $\Pi$ of Eq. (9) must be
continuous across the shock (that defines `some' region of
energy close to the point $M$ on the curve.). 

So far, we discussed the nature of solutions 
when both the saddle type sonic points are present. This is true only 
in an energy range  $\Delta {\cal E}$ spanned by the curve $CMD$ (see, Fig. 3 below). 
Otherwise, the flow
will have only one saddle type sonic point, and it will surely not
have a standing shock wave of the kind we are discussing here (though
shocks by external irradiation which forces the flow to be
subsonic cannot be ruled out). The whole range of $\Delta {\cal E}$
does not allow shock solution, but only a part, in which Rankine-Hugoniot
condition is satisfied, is allowed. This will be shown below.

\begin{figure}
\psfig{figure=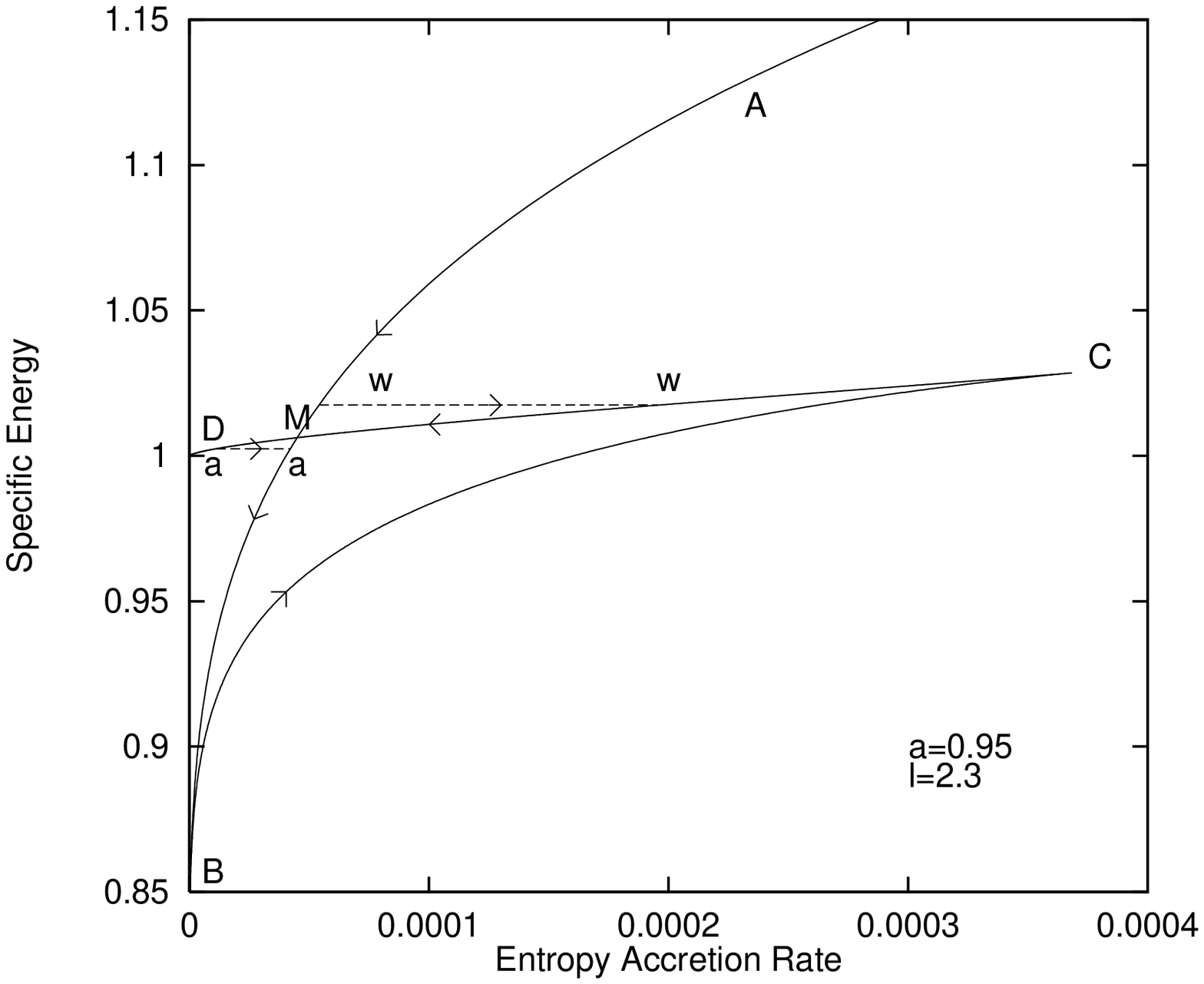,height=8.0truecm,width=8.0truecm,angle=270}
\noindent {\small {\bf Fig. 2:}
Plot of specific energy ${\cal E}$ against the entropy accretion rate
${\dot{\cal M}}$ at all the sonic points of the flow of angular
momentum $l=2.3$ around a Kerr black hole of $a=0.95$. Sections $AMB$
and $CMD$ denote results on inner and outer saddle type sonic points respectively,
and the section $BC$ denotes results of center type sonic point. Horizonal
arrows $aa$, and $ww$ schematically show shock transitions in accretion
and winds respectively. Arrows on the curves show the direction in which
the sonic point is increased.}
\end{figure}

\begin{figure}
\psfig{figure=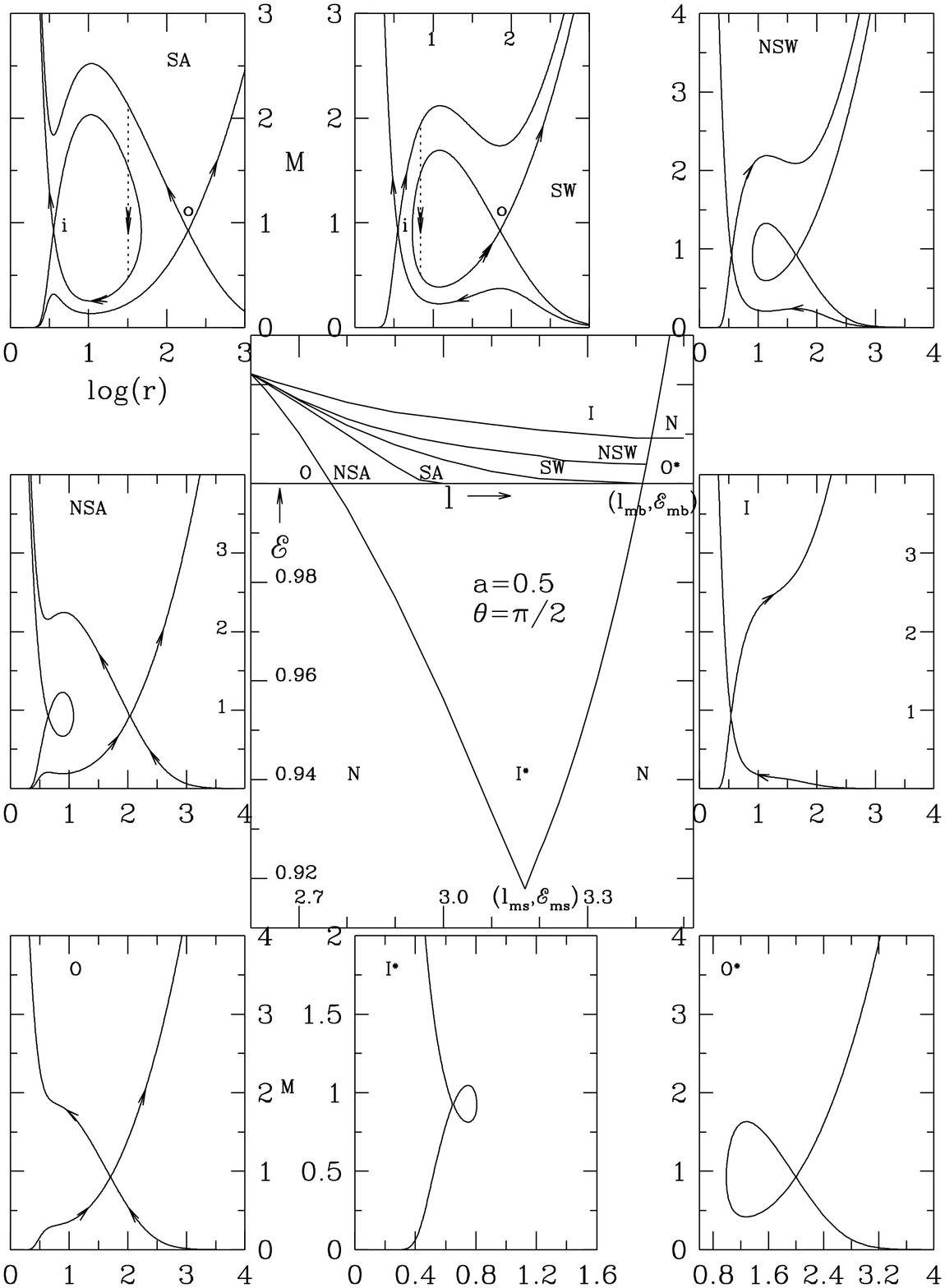,height=9.0truecm,width=9.0truecm,angle=0}
\noindent {\small {\bf Fig. 3:} 
Classification of the parameter space (central box)
in the energy-angular momentum plane in terms of various
topology of the black hole accretion. Eight surrounding boxes
show the solutions from each of the independent regions of the
parameter space. Each small box shows Mach number $M$ against
the logarithmic radial distance $r$. Contours are of constant entropy accretion rate
${\dot{\cal M}}$. Parameters from region N does not form any transonic solution.
See text for details.}
\end{figure}

In Fig. 3 we classified the {\it entire} parameter space
according to the type of inviscid solutions that is prevalent. We choose $a=0.5$.
(For classification of flows in pseudo-Newtonian geometry, see C89, C90ab).
The adiabatic index $\Gamma=4/3$ has been chosen. In the central box,
we divide the parameter space spanned by ($l, {\cal E}$)
into nine regions marked by $N$, $O$, $NSA$, $SA$, $SW$, $NSW$, $I$,
$O^*$, $I^*$. The horizontal line at ${\cal E}=1$ corresponds to the rest
mass of the flow. Surrounding this parameter space, we plot various
solutions (Mach number $M=v_r/a_s$ vs. logarithmic radial distance
where $v_r$ is the radial velocity and $a_s$ is the sound speed) marked
with the same notations (except $N$). Each of these solution topologies has
been drawn using flow parameters from the respective region of the central
box. The accretion solutions have inward pointing arrows and the 
wind solutions have outward pointing arrows.
The crossing points are `X' type or saddle
type sonic points and the contours of circular topology are around `O' type
sonic points. If there are two `X' type sonic points, the inner one is
called the inner sonic point and the outer one is called the outer sonic point.
If there is only one `X' type sonic point
in the entire solution, then the terminology of inner or outer
is used according to whether the sonic point is closer to or farther away from the
black hole. The solutions from the region `O' has only the outer sonic point.
The solutions from the regions $NSA$ and $SA$ have two `X' type sonic points
with the entropy density $S_o$ at the outer sonic point {\it less} than the
entropy density $S_i$ at the inner sonic point. However, flows from $SA$
pass through a standing shock  since the Rankine-Hugoniot
condition is satisfied. The entropy generated at the shock 
$S_i-S_o$ is advected towards the black hole to enable the flow to pass
through the inner sonic point. Rankine-Hugoniot condition is not satisfied
for flows from the region $NSA$. Numerical simulations show
(Ryu, Chakrabarti \& Molteni, 1997) that the flow from this region is very unstable
and exhibit periodic changes in emission properties as the flow
constantly tries to form the shock wave, but fails to do so. The solutions
from the region $SW$ and $NSW$ are very similar to those from
$SA$ and $NSA$. However, $S_o \geq S_i$ in these cases.
Shocks can form only in winds from the region $SW$. The shock condition is not
satisfied in winds from the region $NSW$. This may make the $NSW$ flow
unstable as well. A flow from region $I$ only has the inner sonic
point and thus can form shocks (which require
the presence of two saddle type sonic points) only if the inflow
is already supersonic due to some other physical processes.
Each solution from regions $I^*$ and $O^*$ has two sonic points (one `X' and one `O')
only and neither produces complete and global solution. The region $I^*$
has an inner sonic point but the solution does not extend subsonically
to a large distance. The region $O^*$ has an outer sonic point, but the
solution does not extend supersonically
to the horizon! 
When a significant viscosity is added, the closed topology of $I^*$
opens up as described in C90ab and C96, and the flow may join
with a cool Keplerian disk with ${\cal E} <1$. These special
solutions of viscous transonic flows should not have
centrifugally supported shock waves as they have only
one inner sonic point. However, hot flows deviating from
a Keplerian disk or, sub-Keplerian winds from companions
or, cool flows subsequently energized by magnetic flares
(for instance) will have ${\cal E}>1$ and thus could have
standing or periodically varying shock waves as
discussed above. The post-shock flow radiates most
of the observed hard radiation as shown by
Chakrabarti \& Titarchuk (1995). Note that parameters from
region $N$ does not produce any transonic solution.

In C89 and C90ab, it was found that shock conditions
were satisfied at four locations: $r_{s1},\ r_{s2},\ r_{s3}, \ r_{s4}$,
though $r_{s1}$ and $r_{s4}$ were found to be
not useful for accretion on black holes. Out of $r_{s2}$ and $r_{s3}$,
it was shown that $r_{s3}$ is stable for accretion flow  and $r_{s2}$
is stable for winds  (Chakrabarti \& Molteni, 1993, also see, Nakayama, 1994
and Nobuta \& Hanawa, 1994). 
We plot here only $r_{s3}$ here in $SA$ and $r_{s2}$ in $SW$
solutions. Here, $o$ and $i$ are the outer and inner sonic points respectively. 
In the box containing a solution from $SA$, 
we chose $a=0.5$, $l=3$, ${\cal E}=1.003$. For these parameters,
the eigenvalue of the critical entropy accretion rates at the two saddle type sonic 
points are ${\dot{\cal M}}_i=2.74 \times 10^{-5}$ and
${\dot{\cal M}}_o=1.491 \times 10^{-5}$ respectively. Here, 
${\dot{\cal M}}_o < {\dot{\cal M}}_i $, hence the flow 
through the outer sonic point joins the horizon with infinity (single arrowed curve).
The flow forms a shock and jump onto the branch which passes through
$i$ as shown by double arrows. The stable shock (shown by a vertical dashed
line) is located at $a_3=r_{s3}=32.29$
(in C89 notation). Only this jump, namely, a generation of entropy of amount
${\dot{\cal M}}_i- {\dot{\cal M}}_o$ is allowed
in order that the transonicity of the post-shock flow is guaranteed.
The entropy generated at the shock is advected through the inner sonic point.
The flow is inefficiently cooled, which keeps the energy of the flow
constant. This makes the flow much hotter than a Keplerian disk.
(This is typical of advective disks. See, C89, C96.)
In $SW$ solution, we chose $a=0.5$, $l=3$, ${\cal E}=1.007$. For these parameters,
the eigenvalue of the critical entropy accretion rates at the two saddle type sonic 
points are ${\dot{\cal M}}_i=3.12 \times 10^{-5}$ and
${\dot{\cal M}}_o=5.001 \times 10^{-5}$ respectively. Here, 
${\dot{\cal M}}_i < {\dot{\cal M}}_o $, hence the flow 
through the inner sonic point $i$ joins the horizon 
with infinity (single arrowed curve).
The accretion flow can no longer form a shock. But a wind,
first passing through $i$ can, as shown in double arrows. The stable shock
(shown by a vertical dashed line) is located at $w_2=r_{s2}=6.89$ in this case. 
Only this jump, namely, a generation of entropy of amount
${\dot{\cal M}}_o- {\dot{\cal M}}_i$ is allowed at the shock
in order that it can escape to infinity through the outer
sonic point $O$. This consideration, along with the continuity of
$\Pi$ (Eq. 19) allows one to locate stationary shock waves in a flow.
Note that though the flow has a shock-free solution (passing through 
$o$ for accretion in $SA$ solution and through $i$ for winds in $SW$ 
solution in Fig. 3),
the flow would choose to pass through a shock because the latter solution is of
higher entropy. This fact has been verified through numerical simulations
of accretion and wind flows (Chakrabarti \& Molteni, 1993; Molteni, Lanzafame \& Chakrabarti,
1994). It is to be noted that the angular momenta associated with
solutions which include shocks are not arbitrarily large. Rather, they are
typically less than the marginally stable value $l_{ms}$ as indicated in Fig. 3.

\begin{figure}
\psfig{figure=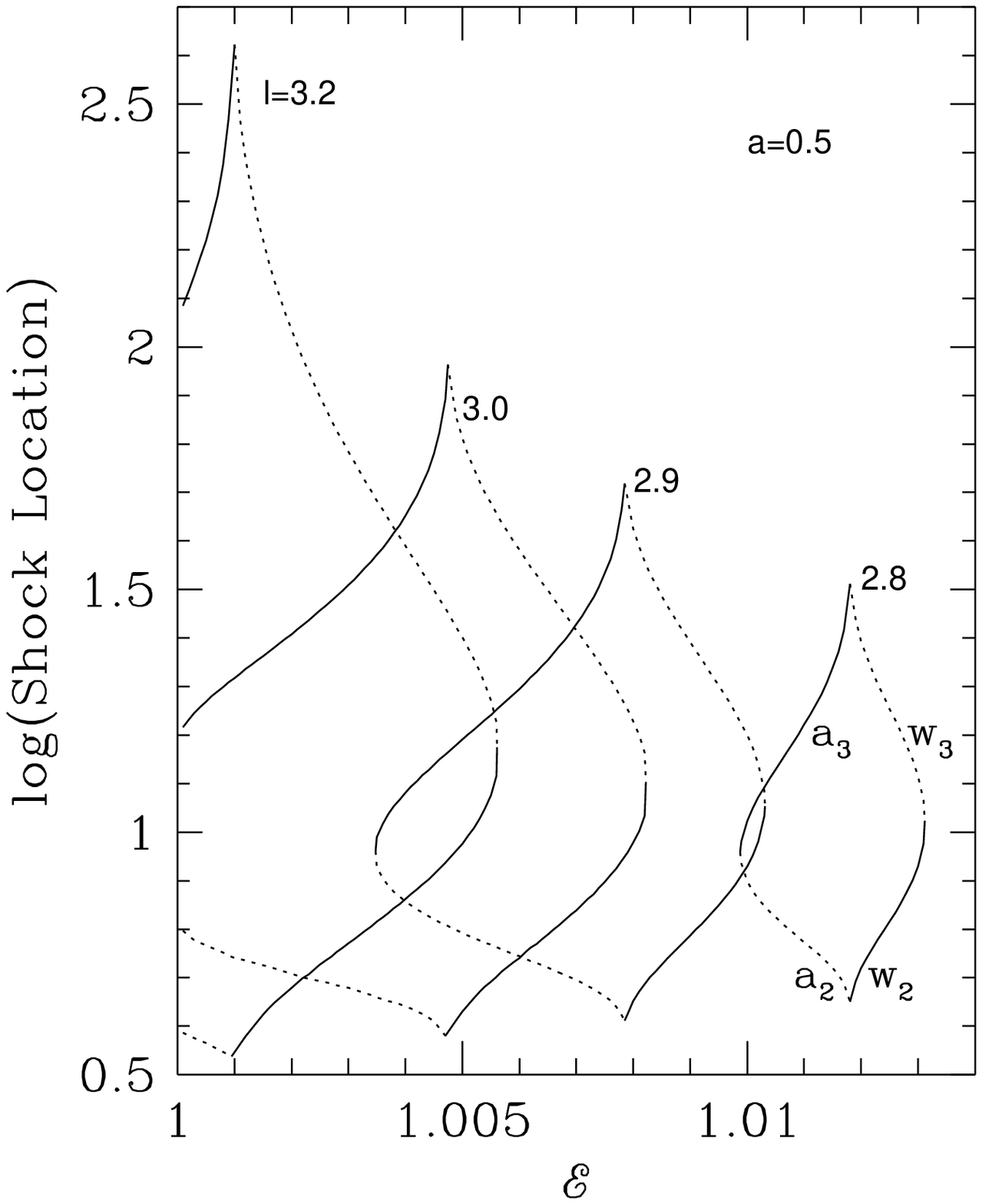,height=10.0truecm,width=10.0truecm,angle=0}
\noindent{\small {\bf Fig. 4:} 
Variation of shock locations  with energy in accretion and winds for various
specific angular momenta $l$ (marked on curves). $a=0.5$ is chosen.
Segments marked $a_3$ and $w_2$ (solid curves) represent stable
shocks in accretion and winds respectively. Other two segments
($a_2$ and $w_3$) represent formal shock locations which are unstable.}
\end{figure}

\begin{figure}
\psfig{figure=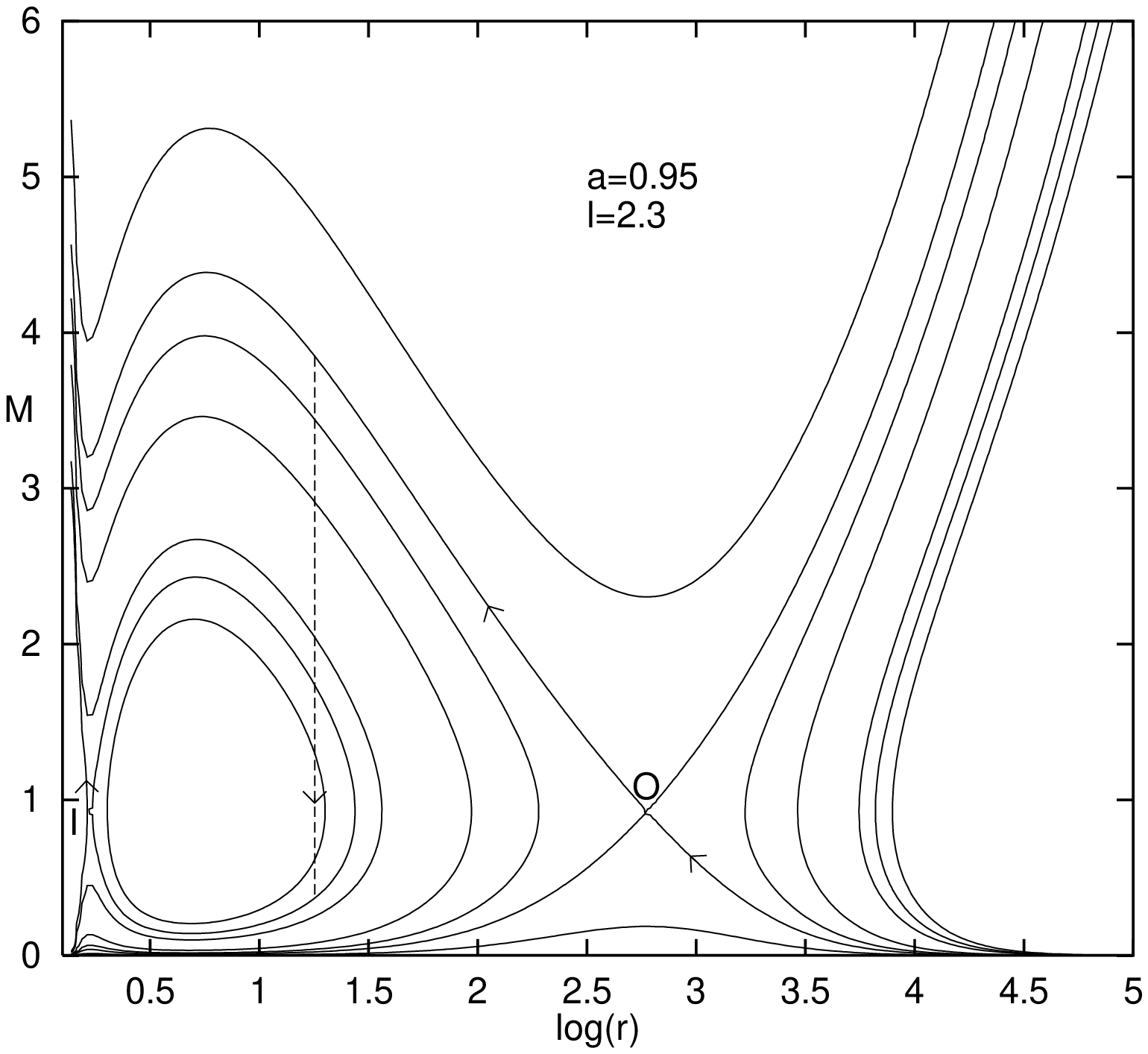,height=8.0truecm,width=8.0truecm,angle=270}
\noindent{\small {\bf Fig. 5a:} Example of shock transition
in a prograde flow.
The parameters are: $a=0.95$, $l=2.3$, ${\cal  E}=1.001$. }
\end{figure}

\begin{figure}
\psfig{figure=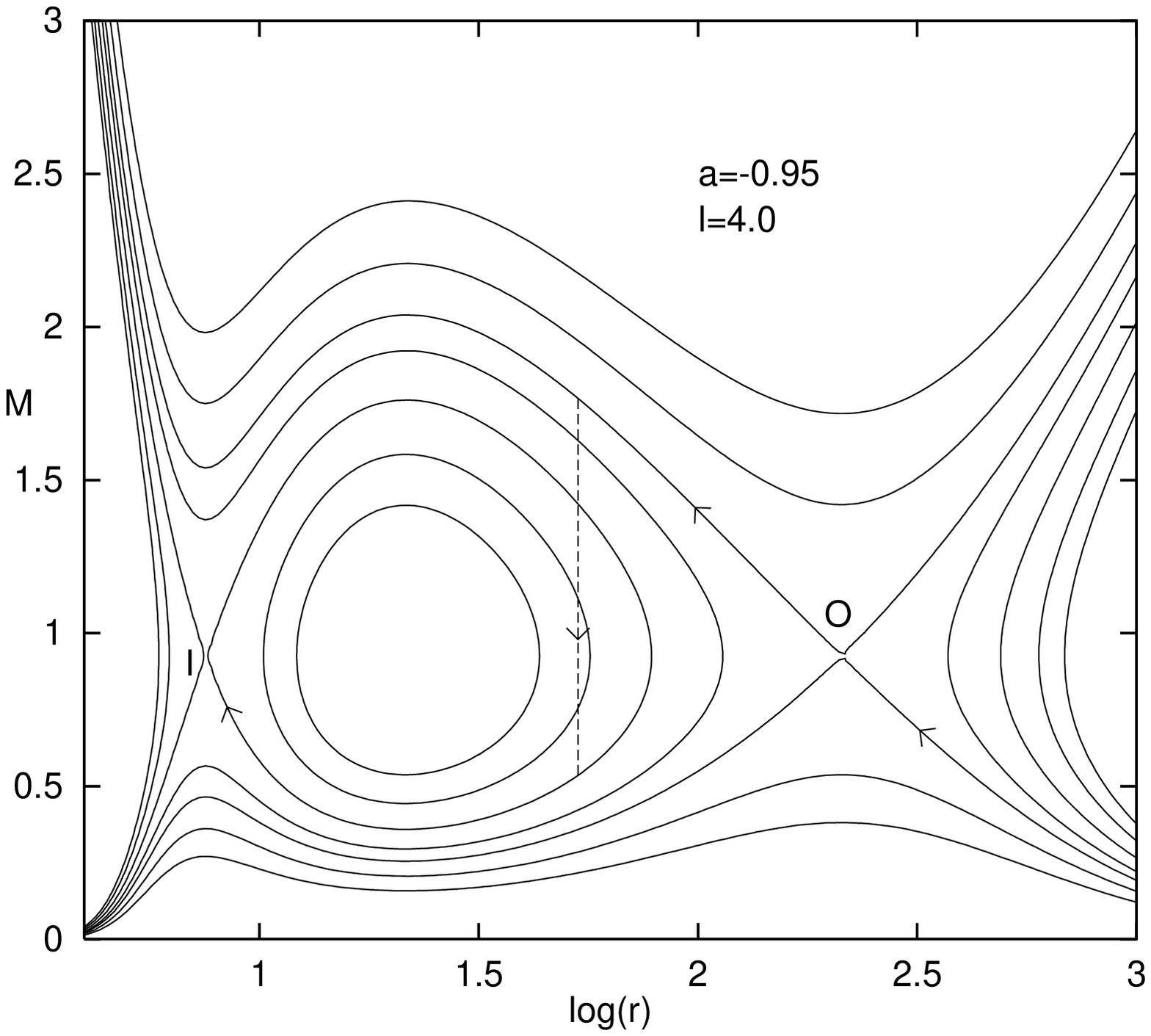,height=8.0truecm,width=8.0truecm,angle=270}
\noindent{\small {\bf Fig. 5b:} Example of shock transition
in a retrograde flow.
The parameters are: $a=-0.95$, $l=4.0$, ${\cal  E}=1.003$. 
 }
\end{figure}

Global solutions which contain shock waves
are not isolated solutions, but are present in a large range of energy and angular momentum. 
In Fig. 4, we show the variation of shock locations as a function of
specific energy ${\cal E}$. Each set of curves, drawn for various specific angular
momentum (marked on the set), consists of four segments: 
two for accretion ($a_2=r_{s2}$ and $a_3=r_{s3}$) and two for 
winds ($w_2=r_{s2}$ and $w_3=r_{s3}$). As discussed above, $a_2$
and $w_3$ (dotted curves) are unstable while $a_3$ and $w_2$ (solid curves) are
stable. Kerr parameter $a=0.5$ is chosen. 
This example shows that stable shocks can form for a very wide class of flows.
For corotating flows, the marginally stable and marginally bound angular momenta 
are $l_{ms}=2.9029$ and $l_{mb}=3.4142$ respectively. Thus the shocks form 
for angular momentum around these values. Since centrifugal barrier becomes
stronger with angular momentum, shocks are located at larger radii 
for higher angular momenta. Another important point to note is that
the shock location increases when the specific 
energy is increased. In a quasi-spherical flow,
with the same input radial velocity and angular momentum, the potential energy
decreases with height (since the gravity becomes weaker), thereby increasing the specific
energy. Thus, shocks in a three-dimensional flow are expected to
bend outwards. Such shocks have indeed been seen in numerical simulations of
Hawley et al. (1984, 1985) and Molteni et al., 1996. Note that the parameters for stable 
shocks in accretion and winds are mutually exclusive. This is a consequence
of the second law of thermodynamics, which requires that ${\dot{\cal M}}_+ \geq 
{\dot{\cal M}}_-$ (C89).

We now compare the nature of shock solutions in prograde and retrograde
flows. In Fig. 5a, we show a shock solution in prograde accretion with
$a=0.95,\ l=2.3, \ {\cal E}=1.001$. The eigenvalues of the 
entropy accretion rates at the outer ($O$) and inner ($I$)
sonic points are, ${\dot{\cal M}}_o=2.973 \times 10^{-06},\ 
{\dot{\cal M}}_i=4.108 \times 10^{-5}$ respectively and the stable shock is
located at $ r_{s3}=18.287$.
In Fig. 5b, we show a shock solution in retrograde accretion with
$a=-0.95,\ l=4.0, \ {\cal E}=1.0025$. The eigenvalues of the 
entropy accretion rates at the outer ($O$) and inner ($I$)
sonic points are, $ {\dot{\cal M}}_o=1.097 \times 10^{-05},
\ {\dot{\cal M}}_i=1.478 \times 10^{-05}$ and $r_{s3}= 53.41$. 
For a retrograde flow, shock locations are generally  higher than that for
a prograde flow. Not shown in the figure is an interesting fact that
the Mach number of the flow remains perfectly finite (typically in the range $5-8$
in our examples) on the horizon.

\section{Global Nature of Shear Tensor}

In presence of viscosity, angular momentum transport is determined 
by the shear tensor. Whereas the computation of actual shear must be done
in conjunction with the global solution of velocity components,
presently we discuss its nature when the viscosity is weak, so that our
inviscid global solution may be used instead. Around a Newtonian star,
shear tensor component due to pure rotational motion can be defined as 
$\sigma_{r\phi}=- r d\Omega/dr$ which is always positive ($ \sim 3/2 \Omega$
for a Keplerian disk), and hence the viscous stress is always negative
(that is why $-\alpha P$ viscosity prescription of Shakura-Sunyaev [1973] was admissible). The
viscous force in the $\phi$ direction, caused by rubbing the adjacent 
layers generates a torque which would carry angular momentum only outwards. 
However, Anderson \& Lemos (1988) 
pointed out that the shear can change sign near the horizon and they 
demonstrated this using velocity profiles of a cold radial 
flow below marginally stable orbit. Below we show that this conclusion 
remains similar for our global solution. It is to be noted that 
this shear is `passive', namely, computed {\it after} the velocity profiles
are known from our inviscid solution. 
On the other hand, this is certainly more accurate than what is obtained by rotational
velocity alone. In presence of small viscosity, the behavior
of our shear is expected to be the same. 
Computations with higher viscosities will be presented elsewhere. 

In Section 2, we presented two definitions of shear tensor, one for the rotationally
dominated flow and the other for the general flow. In Fig. 6, we compare these
two definitions ($\sigma_\phi^r|_{rot}$ and $\sigma_{\phi\pm}^r$) using the velocity profiles from 
our solution. We also plot the gradient of angular velocity $\Omega$ 
(dotted curve). The upper and lower panels are drawn using the 
parameters of Figs. 5a and 5b respectively. Thus they represent typical behaviours
in prograde and retrograde flows respectively. We plot only the
solution branches which pass through the outer sonic point (solid curve
represents $\sigma_{\phi+}^r$ for the supersonic branch, and the long dashed curve represents
$\sigma_{\phi-}^r$ for the subsonic branch). The other 
branches which pass through the inner sonic point
behave in a similar manner. Also shock-free global solutions
show a similar behaviour. The difference in these two cases is due to
significant radial velocity of the flow. We note that while the shear 
stress derived from rotational motion and that for the subsonic branch
derived from general motion ($\sigma_{\phi-}^r$) vanish on the horizon,
the general shear on supersonic branch ($\sigma_{\phi+}^r$, which is relevant
for our work) does not vanish on the horizon.  
This shows that a flow just outside the horizon
will transport angular momentum towards the horizon. Although for any realistic
viscosity, this behaviour is not expected to make any difference in the
angular momentum distribution, and therefore our earlier computations
(e.g. C90ab) which ignore such effect (by assuming $\alpha P$ viscosity which 
is always positive, say) should remain valid.
The reversal of the viscous stress is clearly due to the coupling of the
rotational energy and the gravitational energy of the flow. Here, the `mass' equivalent
of the rotational energy is also attracted by gravity, making the attractive force stronger
than a that of a Newtonian star while the centrifugal force remains the
same. The gravity always wins (giving rise to the well known `pit in the potential'
for any angular momentum). To remain in equilibrium the flow angular
momentum must also rise as it approaches the black hole horizon and to achieve
this the angular momentum must be transported inwards. This is the origin of the
trend of the shear tensor. The non-linearity (coupling among all the energy terms)
which gives rise to this interesting behaviour is the hall-mark of general relativity
as is discussed in detail in Chakrabarti (1993). 

\begin{figure}
\psfig{figure=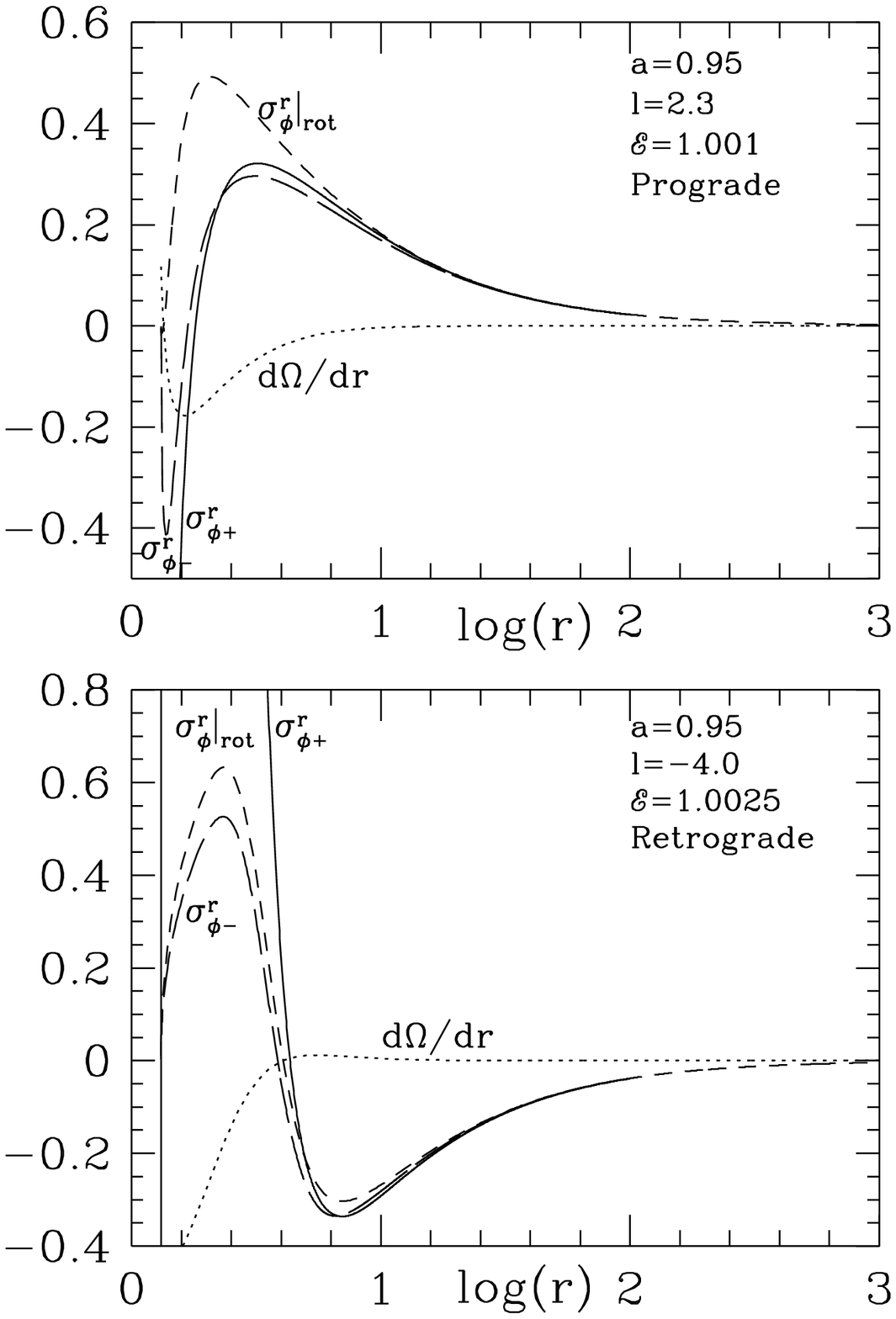,height=10.0truecm,width=10.0truecm,angle=0}
\noindent{\small {\bf Fig. 6:} 
Comparison of rotational shear stress $\sigma_\phi^r|_{rot}$
(short dashed curves) with complete shear stress $\sigma_{\phi+}^r$ (solid)
along the supersonic branch (passing through outer sonic point)
for (a) a prograde flow (upper panel) and  (b) a retrograde flow (lower
panel). Also shown is $d\Omega/dr$ (dotted curves). For
comparison, results of the subsonic branch ($\sigma_{\phi-}^r$) which passes
through the outer sonic point is also shown (long dashed).
Note the change in sign of shear near the horizon. $\sigma_{\phi+}^r$
does not vanish on the horizon, but $\sigma_\phi^r|_{rot}$ and
$\sigma_{\phi-}^r$ do. Here, subscripts `+' and `-' denote supersonic and subsonic branches
respectively. }

\end{figure}

\section{Concluding Remarks}

In this paper we presented the equations governing 
a viscous transonic flow in vertical equilibrium
around a Kerr black hole and obtained the complete set of 
global solutions of them. We showed that for a large region of 
the parameter space, a transonic flow can have a shock
wave as well. We showed that standing shocks can form much closer to the 
black hole than shocks in the Schwarzschild geometry. This causes the 
post-shock region to be much hotter and may thus contribute to very high energy X-rays and 
$\gamma$-rays observed from black hole candidates (Chakrabarti \& Titarchuk, 1995). 
Furthermore, the angular momentum required to have shock waves are also lower,
(typically $l\sim GM_{bh}/c$ for $a\sim 1$) than that required for a Schwarzschild
black hole (typically $l\sim 4GM_{bh}/c$ for $a=0$). This is significant, as even 
with  a small angular momentum the flow can have a stationary shock. It is true that the
region of parameter space which forms a shock is limited, but numerical
simulations (Molteni, Lanzafame \& Chakrabarti, 1994; Molteni, Ryu \& Chakrabarti, 1996)
indicate that flows even outside this range of parameter
contain shock waves, particularly because of the presence of
turbulent pressure whose effects we have not included here. 

Earlier (C89, C90ab, C96) our studies were made using pseudo-Newtonian
potential, where the the inner boundary, namely the horizon was
not satisfactorily described. For instance, the angular velocity
$\Omega=l/r^2$ was not equal to the angular velocity on the
horizon even for a Schwarzschild geometry. Usually this 
was not a problem, since the potential
energy was singular and the rotational energy due to 
any finite angular momentum was negligible compared to it. However,
in our present analysis, such things do not arise. By definition,
$\Omega=u^\phi/u^t$ co-rotates with the horizon and our solution 
continuously or discontinuously extends from the horizon to infinity.

In this paper, we concentrated on the cases 
where the polytropic index $n=3$, i.e., $\Gamma=4/3$. When the
flow is completely gas pressure dominated, the flow may have $\Gamma \sim 5/3$ in
absence of cooling effects.
In this extreme case, the flow will have only one (inner) saddle type sonic
point (C90b) and the shocks can form only if the flow is already supersonic
(e.g., when some wind is accreted). In reality, the flow will have 
intermediate polytropic index, dictated by cooling and heating processes
and the possibility of shock formation will still be significant.
The separation of topologies take place at
around $\Gamma=1.5$ for flows in vertical equilibrium (C96).

We have discussed the behaviour of shear tensor components for our 
transonic solutions when the viscosity is negligible. Viscous stress shows  a
similar trend (namely, developing a maximum) as in a cold, radial flow
(Anderson \& Lemos 1988). However, unlike this case where the reversal
occurs for a slowly rotating black hole, our hot, non-radial
flow shows the trend of reversal even when the Kerr parameter is high.
The conclusions we drew for an inviscid flow cannot change dramatically
when a weak viscosity is present, except that instead of `X' and `O'.
type sonic points, one will have `X' and `spiral' or `nodal' type points
(C90ab, C96). A flow which is cool and strictly Keplerian
must be bound (${\cal E}<1$) and therefore will not have shocks
(Fig. 3) without additional energizing mechanism. The Keplerian disk
will simply become sub-Keplerian due to pressure effects and enter
through the inner sonic point before disappearing into the black hole.
On the other hand, if the Keplerian  disk is hot,
and/or the sub-Keplerian flow originated from a cool Keplerian
disk is re-energized (by magnetic flaring and other effects,
for instance), or, if the sub-Keplerian flows are mixed
with supersonic winds coming from companions
then they can have standing shocks provided viscosity is below
a critical value (C96). In fact, even in an
axisymmetric fully three dimensional flow, in full Kerr
geometry, the shocks form more easily above and below the
equatorial plane (Chakrabarti, in preparation) where the accretion is
presumably highly advective and sub-Keplerian. These considerations were
used by Chakrabarti \& Titarchuk (1995) to prove that the transition of
spectral states of black hole candidates is due to variation of
accretion rates. The analysis of 1100 days of BATSE data of Cyg X-1
(Crary et al, 1996) shows an excellent agreement with such
considerations. In the absence of shocks,
the spectral properties may vary similarly (our conclusions
regarding the production of weak hard tail component
in converging flows are independent of whether shocks form
or not) since viscous effects (or, equivalently, accretion
rates) change the transition radius where Keplerian
disk becomes sub-Keplerian (C90ab; Chakrabarti \& Titarchuk,
1995; C96). However, in this case variations in the hard
and soft components are always correlated which are not observed
in most of the black hole candidates, especially in wind fed systems.
Thus, we believe two independent accretion components, namely, Keplerian
and sub-Keplerian, should be present rather than one single component
becoming sub-Keplerian and producing hard radiations. The most general
solutions of advective disks in pseudo-Newtonian potential discussed
these properties (C96) and we showed in the present paper that the
general conclusions do not change even when the computations are carried out
in full general relativity.

In the present paper, we have written the most general equations
in Kerr geometry which govern hydrodynamics 
of the flow (originally attempted by NT73 in the Keplerian limit) 
by including radial velocity, and correcting the angular momentum 
transport equations with the use of conserved proper angular
momentum and proper shear components. We already showed that the 
relevant shear components behave very differently from that of a flow
around a Newtonian star. In future, we will present solutions of these
equations for highly viscous flows.

The author thanks Kip Thorne and J. Peitz for discussion.

\end{document}